\documentclass{article}

\usepackage{PRIMEarxiv}

\usepackage[utf8]{inputenc} % allow utf-8 input
\usepackage[T1]{fontenc}    % use 8-bit T1 fonts
\usepackage{hyperref}       % hyperlinks
\usepackage{url}            % simple URL typesetting
\usepackage{booktabs}       % professional-quality tables
\usepackage{amsfonts}       % blackboard math symbols
\usepackage{nicefrac}       % compact symbols for 1/2, etc.
\usepackage{microtype}      % microtypography
\usepackage{lipsum}
\usepackage{fancyhdr}       % header
\usepackage{graphicx}       % graphics
\graphicspath{{media/}}     % organize your images and other figures under media/ folder
\usepackage{amsmath,amssymb}
\usepackage[ruled,vlined]{algorithm2e}
\usepackage{subcaption}
\usepackage{caption}
\usepackage{booktabs} % For formal tables
\usepackage{siunitx} % Provides the S column type
\usepackage{amsmath} % For text in math mode
\usepackage{cite}
\usepackage{graphicx}
\usepackage{caption}
\begin{document}
% Update your Headers here
% \fancyhead[LO]{Running Title for Header}
% \fancyhead[RE]{Firstauthor and Secondauthor} % Firstauthor et al. if more than 2 - must use \documentclass[twoside]{article}

%% Title
\title{Advancing Algorithmic Trading: A Multi-Technique Enhancement of Deep Q-Network Models
%%%% Cite as
%%%% Update your official citation here when published 
% \thanks{\textit{\underline{Citation}}: 
% \textbf{Authors. Title. Pages.... DOI:000000/11111.}} 
}

\author{
  Gang Hu \\
  School of Computer Science \\
  Georgia Institute of Technology \\
  Atlanta\\
  \texttt{\ ghu70@gatech.edu} \\
}
\maketitle

\begin{abstract}
This study investigates the enhancement of the traditional Deep Q-Network (DQN) trader model through the integration of cutting-edge techniques such as Prioritized Experience Replay, Regularized Q-Learning, Noisy Networks, Dueling DQN, and Double DQN. Through rigorous empirical testing on a spectrum of financial instruments including BTC/USD and AAPL, the research delineates clear performance improvements over the original model. The augmented DQN trader showcases remarkable gains, with the DQN-vanilla variant achieving an arithmetic return increase from 261\% to 287\% and an enhanced Sharpe Ratio, indicative of better risk-adjusted returns. The innovative use of CNN1D and CNN2D architectures further amplifies returns, highlighting the efficacy of convolutional layers in capturing market dynamics.

The enhanced model's consistency is evident in its application to AAPL stock, where substantial gains are observed. The DQN-pattern variant maintains a stable performance, while the CNN-based models demonstrate their architectural potency through exceptional returns. These results not only eclipse those of the baseline model but also underscore the potential of utilizing convolutional neural networks within financial trading systems.

The study's findings confirm that the application of these sophisticated deep learning techniques within a reinforcement learning framework can significantly improve the performance of automated trading strategies. This performance consistency across various financial instruments underpins the importance of continued innovation in this domain. The research concludes by advocating for future exploration into new reinforcement learning methods and their potential to expand the model's effectiveness across a wider array of financial environments. The abstract encapsulates the core advancements and implications of the study, setting the stage for future developments in the realm of AI-driven financial trading.

\end{abstract}

% keywords can be removed

\section{Introduction}
The quest for profitable investment approaches, tailored to the dynamics of specific financial entities or a portfolio within a particular market, is a cornerstone of financial strategy, more so in light of vast repositories of historical financial data. The emergence of algorithmic trading \cite{chan2021quantitative} has revolutionized market operations, particularly due to the integration of high-frequency trading systems monitored by advanced computing solutions \cite{gomber2015high}. This revolution has sparked an intensive exploration into potent computational models adept at forging successful trading strategies. In this landscape, the deployment of Machine Learning (ML) and Deep Neural Network (DNN) methodologies has been pivotal in advancing the sophistication of trading strategies, encompassing individual asset and portfolio management \cite{Zhang_2020}. The research domain has shown a predilection for adopting genetic programming (GP) and deep reinforcement learning (DRL) for crafting bespoke trading rules that are specific to certain financial assets \cite{taghian2022learning}.

Genetic programming has been a stalwart in the domain, applied with success to distill trading principles for major indices like the S\&P 500 \cite{allen1999using}, and has shown proficiency in generating rules that exploit ephemeral market volatilities \cite{potvin2004generating}. It has also been instrumental in formulating robust rules that withstand market noise, utilizing extensive arrays of technical indicators \cite{chien2010mining}, and in establishing rules underpinned by eminent technical indicators such as the MACD \cite{taghian2021reinforcement}. Nevertheless, the inability of genetic programming to adapt post-deployment has necessitated its combination with reinforcement learning to complement its evolutionary scope.

The remarkable efficacy of Deep Reinforcement Learning (DRL) frameworks, which synergize Deep Neural Networks (DNNs) with reinforcement learning tenets, has been pivotal in shaping contemporary investment strategy development. Modern portfolio management techniques are increasingly reliant on the sophisticated structures of DNNs, augmented by the strategic acumen that DRL methodologies provide. In the realm of stock selection and trading, researchers such as Mehran Taghian, Uta Pigorsch, and others have endeavored to adapt DQN models\cite{taghian2022learning,pigorsch2022high}. Their investigations have revealed that, in comparison with three fundamental strategies—namely, Buy-and-Hold, where an equal-weighted portfolio of all stocks is maintained; Momentum, where stocks with positive average returns over the past five trading days are purchased; and Reversion, where stocks with negative average returns over the same period are acquired—the DQN consistently outperformed in 36 out of 48 experimental setups. Notably, the efficacy of the DQN strategy improved with the enlargement of the investment portfolio, indicating a positive correlation between the number of stock options available and the strategy's performance. Furthermore, the DQN approach exhibited superior performance relative to the benchmark strategies when applied to portfolios comprising stocks with smaller market capitalizations.

In comparison to momentum and reversion strategies, the Deep Q-Network (DQN) strategy has demonstrated superior adaptability to transaction costs. However, despite its various advantages in stock trading, the original DQN model used by researchers is not without its limitations and potential areas for refinement. Firstly, there is the well-documented issue of overestimation in action value estimates inherent in DQN. This overestimation arises from the maximization step in DQN, leading to an overly optimistic assessment of action values, which has been widely observed in both theoretical and empirical studies. Such continual overestimation can cause instability in the learning process and result in suboptimal strategies \cite{van2016deep}.

Beyond the overestimation issue, in standard DQN, Q-values are direct estimates for each state-action pair, without distinguishing between state value and action advantage. This can lead to inefficiencies, as the network must learn the value for each state-action pair independently rather than sharing information about state value. Moreover, standard DQN tends to overestimate action values as it always selects the largest Q-value for updating, which can lead to unstable training dynamics and suboptimal policy. The lack of separation between state value and action advantage in value estimation may also require longer times to converge to the optimal policy, especially in environments with large action spaces. Additionally, the absence of a structured approach to learning could impede the generalization performance of the standard DQN.

Furthermore, the equal treatment mechanism in DQN's experience replay also needs enhancement. This equal treatment can result in a learning process that lacks focus and efficiency, as all experiences are treated with the same importance regardless of their actual contribution to the learning of the value function. This may lead to an overrepresentation of transitions that offer little to no improvement to the agent, thereby slowing the learning process and delaying policy optimization. Additionally, this non-differentiated approach might neglect or under-sample those infrequent but crucial experiences, which could slow the rate of performance enhancement in the long term compared to methods employing prioritized experience replay.

Lastly, while the primary advantage of using a standard linear network within the DQN framework lies in its structural simplicity and ease of implementation, these networks may not be sufficient for capturing the necessary features and decision boundaries for more complex problems. Although they are widely used due to their straightforward design and ease of understanding, and they usually involve fewer parameters and lower computational complexity—resulting in more efficient and faster training—for problems that require exploration, their performance can be unsatisfactory. Moreover, DQN faces issues such as low sample utilization efficiency, high complexity, poor stability, difficulty in tuning, and substantial computational resource consumption.

In order to solve above problems,we initially substituted the conventional DQN with both Double DQN and Duel DQN with the aim of enhancing the performance of the original model and addressing some of the issues previously mentioned. The introduction of Double DQN primarily targets the overestimation bias of action values that can occur with standard DQN. Since DQN utilizes the same network for both action selection and evaluation, it may tend to select overestimated action values. The fundamental concept of Double DQN involves the use of two distinct networks: a behavior network for action selection and a target network for the evaluation of the chosen action's value, with the main goal of diminishing the overestimation bias during action selection. As indicated in the literature \cite{van2016deep}, Double DQN mitigates the overestimation bias found in the original DQN approach, thereby enhancing learning stability and policy performance. Dueling DQN introduces a novel network architecture that more accurately estimates state values and advantage values. Unlike the standard DQN where a neural network outputs the Q-value for each action directly for a given state, Dueling DQN separates into two pathways: one estimating the state-value function (V(s)), which signifies the intrinsic value of a state regardless of the action taken; and another estimating the advantage function (A(s, a)) for each action, reflecting the potential value of choosing a particular action over an average action. These pathways converge towards the end of the network to compute the Q-value for each action. By explicitly distinguishing between state values and action advantages in the network structure, Dueling DQN provides a more granular approach to Q-value estimation, proving more efficient and effective for numerous decision-making tasks \cite{wang2016dueling}. Following the optimization of the model with Dueling DQN, further enhancement of the DQN network model is possible by employing Noisy Networks in place of the conventional linear layers within the original DQN network. Pertaining to exploration, Noisy Networks incorporate noise into the network's parameters to encourage exploration in a stochastic manner. This noise can propel the agent to try new, potentially unassessed actions, possibly accelerating learning in some scenarios. In contrast, conventional linear networks often rely on external strategies for guidance in exploration, such as the $\epsilon$-greedy policy, which may not be as flexible or effective as intrinsic exploration mechanisms. Generally speaking, conventional linear layers hold advantages in simplicity, parameter count, and computational efficiency, yet may not perform as well as Noisy Networks with integrated noise in tasks requiring complex exploratory behaviors \cite{fortunato2017noisy}. Furthermore, the author has incorporated Regularized Q-Learning into his model, mainly by introducing an L2 regularization term into the loss function, typically to mitigate the risk of overfitting. L2 regularization, often referred to as weight decay, serves to penalize the squared magnitude of the weight parameters. To address issues inherent in standard DQN when utilizing experience replay, the author also employed Prioritized Experience Replay (PER) which enhances the training process of reinforcement learning agents by assigning a higher replay probability to experiences with a greater potential for learning, indicated by a substantial temporal difference error (TD-error). This method accelerates the learning rate by concentrating on experiences that are more likely to significantly impact the agent's learning progress. It improves sample efficiency by ensuring that critical experiences are replayed more frequently, thus making more effective use of the available data. Moreover, PER facilitates the agent's faster correction of value estimates by prioritizing experiences with higher prediction errors, which leads to a quicker convergence of the value function. The dynamic adaptability of PER ensures that the importance of experiences is adjusted over time, corresponding with the evolving learning stage of the agent.

Conversely, standard DQN is challenged by issues that PER aims to resolve. It treats all experiences as equally important, potentially causing inefficiencies in learning as certain experiences may contribute more significantly to learning progress than others. This uniform treatment may decelerate convergence, as it may not prioritize the most informative samples swiftly enough. Additionally, standard DQN may not efficiently explore rare but essential experiences, as these could be overwhelmed in the replay buffer by more common occurrences, leading to a homogenized and potentially less optimal exploration strategy. PER addresses these issues by providing more replay opportunities to experiences with high TD-errors, promoting faster and more effective learning \cite{schaul2015prioritized}.

In this study, we have undertaken the following enhancements and implementations:

\begin{itemize}
  \item Replaced the conventional DQN with Double DQN and Duel DQN to address overestimation biases and improve policy performance, as documented by \cite{van2016deep}.
  \item Introduced Dueling DQN architecture for finer estimation of state values and advantage functions, leading to more efficient decision-making as outlined by \cite{wang2016dueling}.
  \item Utilized Noisy Networks for replacing linear layers in DQN to promote exploration and learning acceleration, further discussed in \cite{fortunato2017noisy}.
  \item Incorporated Regularized Q-Learning to reduce overfitting risks through L2 regularization, known as weight decay\cite{farahmand2011regularization}.
  \item Employed Prioritized Experience Replay (PER) to enhance learning by focusing on critical experiences with high TD-error, thereby increasing sample efficiency and speeding up value function convergence \cite{schaul2015prioritized}.
\end{itemize}

Conventional DQN's limitations are mitigated by these enhancements:
\begin{itemize}
  \item PER addresses the inefficiencies of standard DQN by prioritizing experiences based on their learning potential rather than treating all experiences equally.
  \item By focusing on informative experiences, PER accelerates learning and ensures that rare but critical experiences receive adequate attention in the replay buffer.
\end{itemize}

\section{Related work}
Deep Q-Networks (DQN) signify a paradigm shift in reinforcement learning by integrating Q-learning with deep neural networks to manage high-dimensional input spaces effectively. The innovation of DQN lies in its use of a neural network to approximate the Q-values, allowing generalization across a wide range of state-action pairs without exhaustive exposure to every potential combination. This integration enables the DQN to handle complex inputs such as image data from video games. Core to the DQN framework are its Q-network, which predicts Q-values for all possible actions given a state; its experience replay mechanism, which facilitates the breaking of correlation between sequential observations and enhances data utilization; and its target network, a clone of the Q-network with more infrequent updates, contributing to training stability \cite{mnih2015human}.

Subsequent variants of DQN have been developed to address specific limitations and enhance the algorithm's performance. Double DQN (DDQN) mitigates the overestimation bias inherent in the original DQN by decoupling action selection from action evaluation, employing two distinct networks for these tasks \cite{van2016deep}. Dueling DQN modifies the network architecture to provide separate estimations of state values and action advantages, which are then combined for Q-value determination, offering a nuanced understanding of state value independent of the action effect \cite{schaul2015prioritized}. Prioritized Experience Replay (PER) DQN introduces a non-uniform sampling strategy for the replay buffer, where experiences are weighted by their temporal-difference error, prioritizing those more likely to affect the learning process \cite{schaul2015prioritized}. Noisy DQN incorporates parameterized noise into the network to encourage exploration by introducing stochasticity in action selection \cite{fortunato2017noisy}. Lastly, Regularized DQN enhances the stability and generalization of learning by integrating regularization terms into the loss function, which can also promote exploration \cite{ziebart2010modeling}.

These advancements have propelled DQN variants to the forefront of applications across diverse fields. In video game environments, they achieve superhuman performance by interpreting complex sensory inputs and executing strategic decisions \cite{mnih2015human}. The robotics domain benefits from DQN in tasks such as object manipulation and autonomous navigation, where the algorithm learns through iterative interaction with the environment \cite{jager2022first_page}. in healthcare, DQNs promise to optimize chronic disease management by simulating long-term treatment outcomes \cite{komorowski2018artificial}. The universal applicability of DQN and its derivatives underscores their capability to learn from complex and high-dimensional datasets, making intricate decision-making processes computationally tractable. Furthermore, Financial applications have seen DQN variants employed to deduce optimal trading strategies, capitalizing on the algorithm's predictive power in dynamic systems \cite{bengio2018perspectives}.

Deep Reinforcement Learning (DRL) is increasingly adopted in the realms of financial trading and portfolio optimization. Scholarly work in this domain predominantly focuses on asset trading. Notable investigations pertaining to solitary asset trading include the work by Huang\cite{huang2018financial}, which validates DRL's efficacy in currency pair trading. Employing recurrent neural network structures within DRL frameworks, Huang highlights that a smaller replay memory footprint, contrary to conventional reinforcement learning practices, may yield superior outcomes. Furthermore, an intriguing discovery is that augmented transaction costs do not inherently impede trading efficacy, suggesting that the resultant strategies exhibit enhanced robustness. The author's methodology surpasses established benchmarks, namely the buy-and-hold and sell-and-hold strategies, while also delving into the trading agents' behaviors which tend to sustain win rates around 60\% and marginal profit averages.

Consistent with these observations, our research corroborates the adaptability of agents in navigating elevated transaction costs and outstripping active trading benchmarks, as well as surpassing passive investment strategies in terms of aggregate returns. Mnih et al.'s Deep Q-learning algorithm is widely recognized for its applicability in automated trading of financial assets\cite{mnih2015human}. Théate and Ernst evaluated DQN-based agents across a spectrum of 30 equities and indices, with the agents devising strategies that, on average, eclipsed the performance of a static buy-and-hold strategy\cite{theate2021application}. Their analysis spans an initial training period of six years, followed by a two-year performance assessment. While their agents frequently vacillated between passive and mean-reversion strategies, they fell short of a passive investment benchmark for certain assets. Our findings suggest that the integration and modulation of multiple investment strategies are imperative, as our agents are benchmarked against both passive and active strategies, with neither emerging as conclusively superior in our proposed portfolio configurations.

Li et al. scrutinize various incarnations of Deep Q-learning, endorsing the vanilla version for optimal performance on a selection of U.S. stocks\cite{li2020application}. Zhang et al.\cite{Zhang_2020} extend the comparative analysis of DRL algorithms in both discrete and continuous action spaces, particularly for futures contract trading across diverse asset classes encompassing a pool of 50 contracts. Their findings commend Deep Q-learning's superior performance and underscore the algorithms' robustness against varying levels of transaction costs. In light of these findings, we have adopted the vanilla Deep Q-learning algorithm\cite{mnih2015human} as our fundamental algorithmic framework.

With regard to neural network architectural choices, we draw inspiration from Taghian et al., who compare various feature extraction network architectures tailored to DRL applications in financial trading\cite{taghian2022learning}. Their empirical evidence suggests that a rudimentary multi-layer perceptron architecture, underpinned by the Deep Q-learning algorithm, delivers the most commendable results. This architecture particularly excels when leveraging raw price data time series (including open, high, low, and close prices) as input, which, in their comparative analysis, outperformed hand-crafted time series features such as candlestick patterns. Given the marginal predictive utility of hand-crafted features, our methodology pivots to simplifying the input vector through the utilization of multiple moving averages in conjunction with specific stock data from quarterly reports to diminish input dimensionality.

In an endeavor to optimize deep reinforcement learning (DRL) for financial asset trading, a confluence of progressive techniques has been employed to refine the standard Deep Q-Networks (DQN) framework. The integrative enhancement leveraged through Prioritized Experience Replay (PER), Regularized Q-Learning, Noisy Networks, Dueling DQN, and Double DQN is tailored to address the multifaceted nature of financial markets and surmount the inherent constraints of the original DQN.

The adoption of \textit{Prioritized Experience Replay (PER)} augments the learning efficacy by favoring the sampling of transitions with pronounced prediction errors, thereby allocating greater learning capacity to the most informative experiences \cite{schaul2015prioritized}. This is particularly salient in financial domains where anomalous market events, despite their rarity, bear significant impact.

Incorporating \textit{Regularized Q-Learning} instills a regularization component into the loss function, attenuating overfitting to extant market conditions and endorsing a more resilient policy evaluation \cite{farahmand2011regularization}. This fortifies the strategy against market volatilities and unpredicted scenarios, an indispensable attribute for financial trading algorithms.

\textit{Noisy Networks} are entwined within the network architecture to promote exploration through stochastic perturbations to the weights, thus driving the discovery of novel and potentially lucrative trading policies in the dynamic financial market milieu \cite{fortunato2017noisy}.

The \textit{Dueling DQN} framework bifurcates the assessment of state values from the advantages of particular actions, enabling a granular appraisal of the value of each action in relation to others \cite{wang2016dueling}. Such a distinction is pivotal in discerning the inherent value of a financial state vis-à-vis the relative benefit of a specific trading decision.

Lastly, the \textit{Double DQN (DDQN)} mitigates the overestimation bias of Q-values endemic to the traditional DQN by decoupling action selection from action evaluation \cite{van2016deep}. This engenders a more faithful and steady valuation of actions, a critical factor for navigating the fluctuant realms of financial markets.

Together, these methodological augmentations coalesce to cultivate a DRL model that is better equipped to navigate the complexities and uncertainties of financial asset trading, surpassing the original DQN in learning velocity, strategic stability, and ensuring a judicious balance between exploration and exploitation. These attributes coadunate to potentially forge more efficacious and profitable trading strategies.

\section{Methodology}
\subsection{Trading rules learned by Deep Q-network agent}

A model inspired by the DQN framework is introduced for feature-rich time-series analysis to infer robust trading signals for financial assets. Figure 1 delineates the model's architecture, comprising two primary components: feature extraction and decision making.

\
\begin{figure}[!h]
\centering
\includegraphics[width=\linewidth]{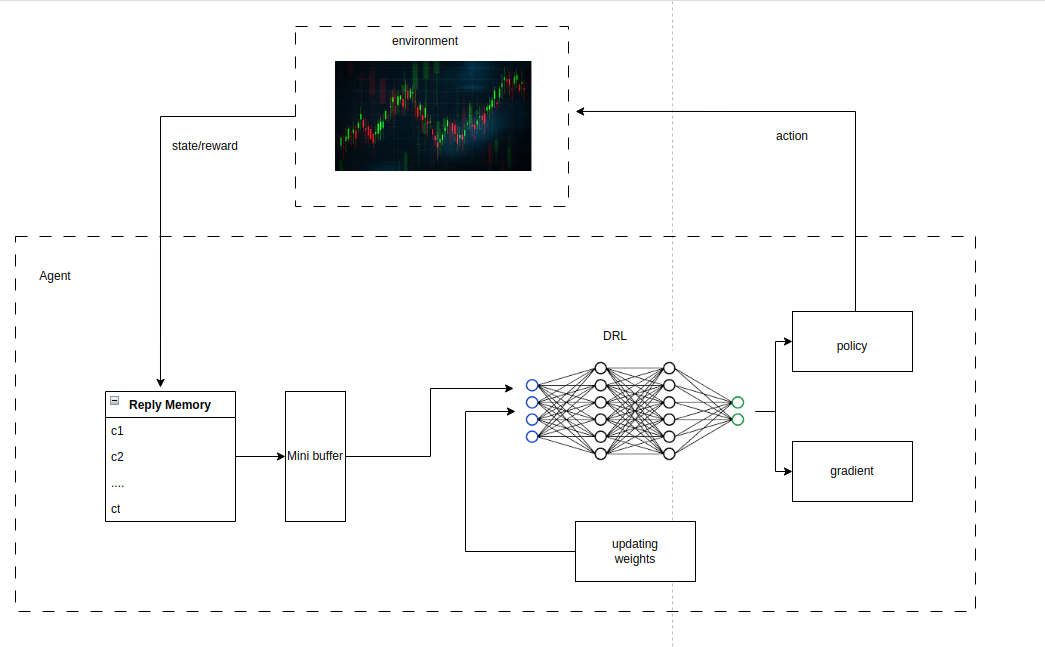}
\caption{{\bf DQN trading Structure.}
In the proposed architecture, the environment provides the state at every discrete time interval. The agent, in response, performs an action contingent on the current state and subsequently receives a reward as well as the subsequent state. This sequence of events, characterized as tuples of (CurrentState, Action, Reward, NextState)}
\label{fig1}
\label{fig:dqn-architecture}
\end{figure}

\subsection{Candlestick Pattern Recognition and Trading Strategy Formulation}

The design of our enhanced trading model hinges on two interconnected modules: the Pattern Recognition Module and the Strategy Formulation Module. Initially, the Pattern Recognition Module meticulously parses the raw Open-High-Low-Close (OHLC) data to craft a nuanced feature vector that distills the salient traits of candlestick formations. This intricate vector, which epitomizes the distinctive patterns observed, is consequently channeled to the Strategy Formulation Module.

The Strategy Formulation Module, empowered by a sophisticated Deep Q-Network (DQN) infrastructure, leverages the extracted features to divine the most advantageous trading action for the next time interval. This module is adept at assimilating both the incoming state vector and the concurrent reward distributed by the market environment, thus facilitating informed and agile decision-making. The ingenuity of this system is encapsulated in its utilization of an advanced neural network for the determination of the Q function, represented by the network weights $\Theta$. This strategic shift from elementary tabular methods to a more profound neural network approach allows for the articulation of potent decision-making rules, especially in the context of rare or complex candlestick configurations that demand precise and astute trading signals.

Rooted in the pioneering frameworks of deep reinforcement learning, our Strategy Formulation Module refines the target Q function through a network of deep neural connections, symbolized by the weights $\Theta$. The network's training regimen is meticulously orchestrated via the Bellman equation, ensuring a systematic approach to the continual enhancement of the model's decision-making acumen.

\begin{eqnarray}
\label{eq:loss_function}
L(\Theta) = \mathbb{E}\left[\left(Q_{\Theta}(S, A) - \hat{Q}_{\Theta}(S, A)\right)^2\right],
\end{eqnarray}

where $\hat{Q}_{\Theta}(S, A)$ is estimated as per the subsequent equation:

\begin{eqnarray}
\label{eq:q_function}
\hat{Q}_{\Theta}(S_t, A_t) = R_t + \gamma \max_{\hat{A}} Q_{\Theta}(S_{t+1}, \hat{A}).
\end{eqnarray}

A separate iteration-frozen network copy aids in stabilizing the training phase, ensuring the static nature of weights within each iteration. The network comprises three fully connected layers, imparting transformation from the input space to latent spaces and ultimately to a tri-dimensional action space. Batch normalization intercedes between layers to maintain stability and mitigate overfitting, culminating in a Softmax layer that probabilistically maps the actions.

\subsection{Signal Processing and Representation Network}

This network, intricately coupled with the strategy formulation module, leverages the propagated errors from strategic decisions to enhance its signal processing capabilities. Through rigorous testing of various architectural paradigms, including Convolutional Neural Networks (CNNs) and Gated Recurrent Units (GRUs), the network is meticulously calibrated to identify the most effective configuration for generating trading signals. Notably, given the concise nature of the input sequences, more streamlined models have emerged as the frontrunners in terms of performance. 

The fusion of the processed signals with the current market trends results in a robust input for the subsequent strategic decision-making module, ensuring a well-rounded representation. Exploratory analysis into varying input structures has resulted in the development of three distinct methods to bolster model performance. The first method employs a binary vector to encapsulate recognized candlestick patterns, providing a clear, discrete representation of the market's historical behavior. The second approach quantifies the relative proportions of a candlestick's elements, as outlined by the following relations:

This comprehensive approach to signal processing and representation sets the stage for a more nuanced and informed trading strategy, tailored to the intricacies of the financial market's fluctuating dynamics.

\begin{eqnarray}
\label{eq:candlestick_components}
\mathrm{upper} = \frac{p_h - \max(p_c, p_o)}{p_h - p_l}, \quad
\mathrm{lower} = \frac{\min(p_c, p_o) - p_l}{p_h - p_l}, \quad
\mathrm{body} = \frac{|p_c - p_o|}{p_h - p_l},
\end{eqnarray}

where $p$ denotes price, with subscripts $h$, $l$, $o$, and $c$ representing high, low, open, and close, respectively. The third method relies on raw OHLC values, entrusting the feature extraction module with discerning an effective representation.

\subsection{trading algorithm}

\begin{algorithm}[H]
\DontPrintSemicolon
\SetAlgoLined
\caption{Enhanced Deep Q-Learning Algorithm with Advanced Techniques}

Initialize replay memory \( D \) to capacity \( N \) with prioritization scheme\;
Initialize action-value function \( Q \) with random weights \( \theta \)\;
Initialize target action-value function \( \hat{Q} \) with weights \( \theta^- = \theta \)\;
Initialize the parameters for the Noisy Layers \( \sigma \) for exploration\;
\For{\( \text{episode} = 1 \) \KwTo \( M \)}{
    Initialize sequence \( s_1 \) and preprocessed sequence \( \phi_1 = \phi(s_1) \)\;
    \For{\( t = 1 \) \KwTo \( T \)}{
        Select action \( a_t \) using policy derived from \( Q \) with Noisy Layers (exploration)\;
        Execute action \( a_t \) and observe reward \( r_t \) and state \( s_{t+1} \)\;
        Set \( s_{t+1} = s_t, a_t \) and preprocess \( \phi_{t+1} = \phi(s_{t+1}) \)\;
        Store transition \( (\phi_t, a_t, r_t, \phi_{t+1}) \) in \( D \) with priority level\;
        Sample random mini-batch of transitions \( (\phi_j , a_j , r_j , \phi_{j+1}) \) from \( D \) with probability proportional to priority\;
        \eIf{episode terminates at step \( j+1 \)}{
            Set \( y_j = r_j \)\;
        }{
            Set \( y_j = r_j + \gamma \max_{a'} \hat{Q}(\phi_{j+1}, a'; \theta^-) \) using Double DQN estimation\;
        }
        Perform a gradient descent step on \( (y_j - Q(\phi_j , a_j ; \theta))^2 \) with respect to the network parameters \( \theta \), with regularization from Regularized Q-Learning\;
        Update the priorities of sampled transitions based on the TD error\;
        Every \( C \) steps reset \( \hat{Q} \) to \( Q \) and recalculate \( \sigma \) for Noisy Layers\;
        Use Dueling DQN architecture to separately estimate state value \( V(s) \) and advantages \( A(s, a) \) for action value computation\;
    }
}
\end{algorithm}

\subsection{Integration of Advanced Deep Q-Learning Techniques}

The proposed algorithm incorporates several sophisticated mechanisms, each aimed at augmenting distinct aspects of the reinforcement learning process. These enhancements facilitate a more efficient exploration strategy, improve convergence stability, and provide a refined estimation of the value functions.

\textbf{Prioritized Experience Replay:}
The conventional experience replay mechanism is augmented with a prioritization scheme. Each stored transition in the replay memory is assigned a priority level, typically predicated on the temporal-difference (TD) error. This prioritization affects the probability of sampling specific transitions, thereby aligning the learning process with the most salient experiences \cite{schaul2015prioritized}.

\textbf{Regularized Q-Learning:}
Regularization techniques are integrated into the loss function during gradient descent optimization. By imposing a penalty on the magnitude of network parameters, the approach mitigates the risk of overfitting, thereby enhancing the generalization capabilities of the learned policy \cite{farahmand2011regularization}.

\textbf{Noisy Networks:}
To address the challenge of exploration, the \( \epsilon \)-greedy policy is supplanted by Noisy Networks. This approach introduces parametric noise into the network weights, thereby inducing a state-dependent exploration capability. The noise parameters are recalibrated at the culmination of each update cycle, ensuring a dynamic balance between exploration and exploitation \cite{fortunato2017noisy}.

\textbf{Dueling DQN:}
The Dueling DQN architecture decomposes the Q-network into two distinct streams: one that estimates the state value function \( V(s) \), and another that computes the advantage for each action \( A(s, a) \). The combination of these streams provides a more nuanced estimation of the action values, particularly in states where the action choice is inconsequential \cite{wang2016dueling}.

\textbf{Double DQN:}
Overestimation of Q-values is a known issue within Q-Learning algorithms. The Double DQN framework addresses this by decoupling the selection and evaluation of actions. This is achieved by employing two sets of weights: one for the policy network (\( \theta \)) and another for a target network (\( \theta^- \)), ensuring unbiased value estimation during the update phase \cite{van2016deep}.

Each of these components is intricately woven into the fabric of the enhanced Deep Q-Learning Algorithm, engendering a synergistic effect that significantly bolsters the learning efficacy.

\subsection{Implementation Details}

The implementation of our deep reinforcement learning algorithm incorporates a neural network architecture synergizing the strengths of Dueling Deep Q-Network (Dueling DQN), Double Q-learning, and Noisy Networks for exploration. Additionally, we refine the learning process with a Regularized Q-learning approach and utilize a Prioritized Experience Replay mechanism for efficient memory utilization.

\textbf{Network Architecture}
Our network, referred to as \texttt{Q\_Network}, consists of a multi-layered structure. The initial layers are responsible for feature extraction, composed of a series of NoisyLinear modules interleaved with Batch Normalization. The NoisyLinear module is a pivotal component of our network design, introducing stochasticity to the learning process and facilitating exploration. Following the feature extraction, the network bifurcates into two streams as per the Dueling DQN architecture: one estimating the state value function and the other computing the advantage for each action. This bifurcation allows for the independent approximation of state and action values, culminating in a combined output that yields the Q-value for each action given the current state.

\textbf{Training Process}
Training proceeds with mini-batch updates, where a batch of experiences is sampled from a Prioritized Experience Replay memory. In line with the Double Q-learning paradigm, the policy network and a periodically updated target network are utilized to decouple the selection and evaluation of actions, mitigating overoptimistic value estimates. The loss function employed is a combination of the temporal difference error, smoothed by an L1 loss, and an L2 regularization term. The regularization is weighted by a factor to control its influence, ensuring the network does not overfit to the observed states and maintains generalization across the state space.

\textbf{Prioritized Experience Replay}
The Prioritized Experience Replay memory is implemented through a SumTree data structure, enabling efficient sampling of experiences based on their priority level, which is proportional to the magnitude of their temporal difference error. This prioritization allows the algorithm to focus on learning from transitions that have a higher expected learning utility.

\textbf{Hyperparameter Selection}
In our exploration of Deep Q-Networks (DQNs) for trading within the highly volatile domains of stock and Bitcoin markets, a deliberate choice was made to implement smaller hyperparameters, specifically in terms of the discount factor ($\gamma$), batch size, replay memory size, and window size. Empirical observations have indicated that such configurations enable the trading agent to place a stronger emphasis on immediate rewards, a beneficial trait given the rapid and unpredictable changes in these financial environments. A smaller batch size fosters more stochastic learning and prevents overfitting to recent price movements, while a reduced replay memory size ensures that the agent's strategy is informed by recent, more relevant market conditions. Furthermore, a narrower window size allows the algorithm to respond more agilely to new market information, thus maintaining the relevance of its trading decisions. Theoretically, this inclination towards smaller hyperparameters corresponds to a more adaptable and responsive learning policy, better suited for the stochastic and non-stationary nature of financial markets. The adoption of these parameters has been substantiated by superior performance in terms of cumulative returns and reduced drawdowns, substantiating their efficacy in the context of our DQN trading models.

\textbf{Regularization Techniques}
To further enhance the robustness of our approach, we introduce a regularization scheme applied to the loss function during training. This regularization not only prevents overfitting but also stabilizes the learning updates, contributing to the overall stability of the training process.

Through these comprehensive implementation details, our network is equipped to handle the complexities of learning optimal policies in challenging environments, demonstrating significant improvements over standard Q-learning approaches.

\section{Experiments}
\subsection{Datasets}
The methodologies posited in this study have been applied to an array of real-world financial datasets that encompass equities, foreign exchange pairings, and digital assets. For benchmarking the model's effectiveness, the same datasets as those employed by \cite{taghian2021reinforcement} were utilized, in addition to historical data pertaining to AAPL, GOOGL, and KSS. For the cryptocurrency segment, the BTC/USD exchange rate was selected, representative of Bitcoin's valuation. The datasets referenced herein are accessible through Yahoo Finance and Google Finance platforms. Consistent with the specifications delineated in Table 1, all datasets were formatted into daily candlestick charts. As depicted in Figure 2, each asset's price history is bifurcated into two distinct segments.

\begin{table}[!ht]
\centering
\captionsetup{justification=centering} % This centers the caption
\caption{{\bf Data used along with train-test split dates.}}
\begin{tabular}{|l|l|l|l|}
\hline
\bf Data & \bf Begin Date & \bf Split Point & \bf End Date \\ \hline
GOOGL & 2010/01/01 & 2018/01/01 & 2020/08/24 \\ \hline
AAPL & 2010/01/01 & 2018/01/01 & 2020/08/24 \\ \hline
KSS & 1999/01/01 & 2018/01/01 & 2020/08/24 \\ \hline
BTC-USD & 2014/09/17 & 2018/01/01 & 2020/08/26 \\ \hline
\end{tabular}
\label{table1}
\end{table}

\begin{figure}[!h]
\centering

\begin{subfigure}[b]{0.49\textwidth}
    \includegraphics[width=\linewidth]{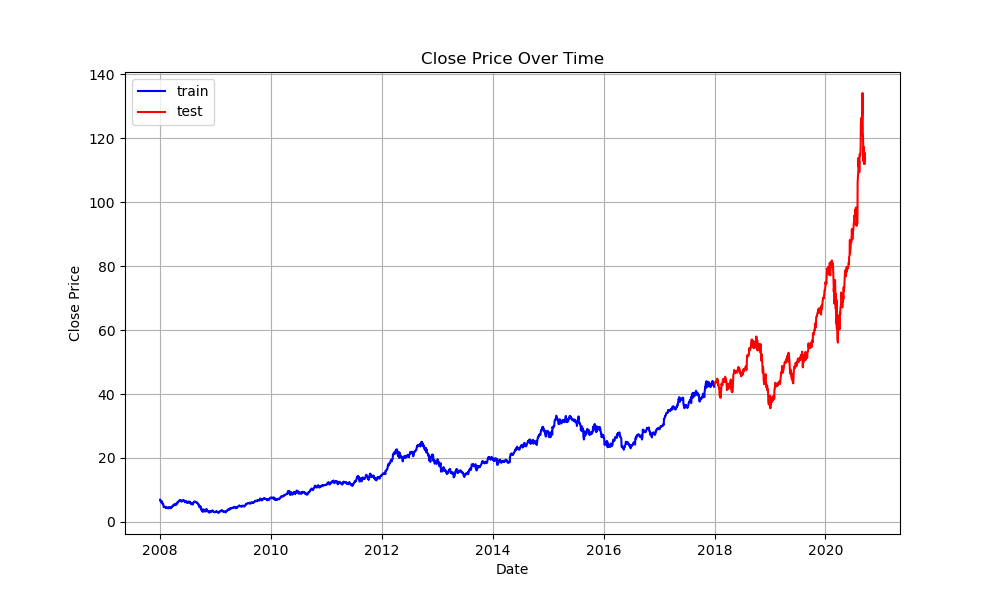}
    \caption{Price History of AAPL}
    \label{fig:aapl}
\end{subfigure}
\hfill % this command puts space between the left and right figures
\begin{subfigure}[b]{0.49\textwidth}
    \includegraphics[width=\linewidth]{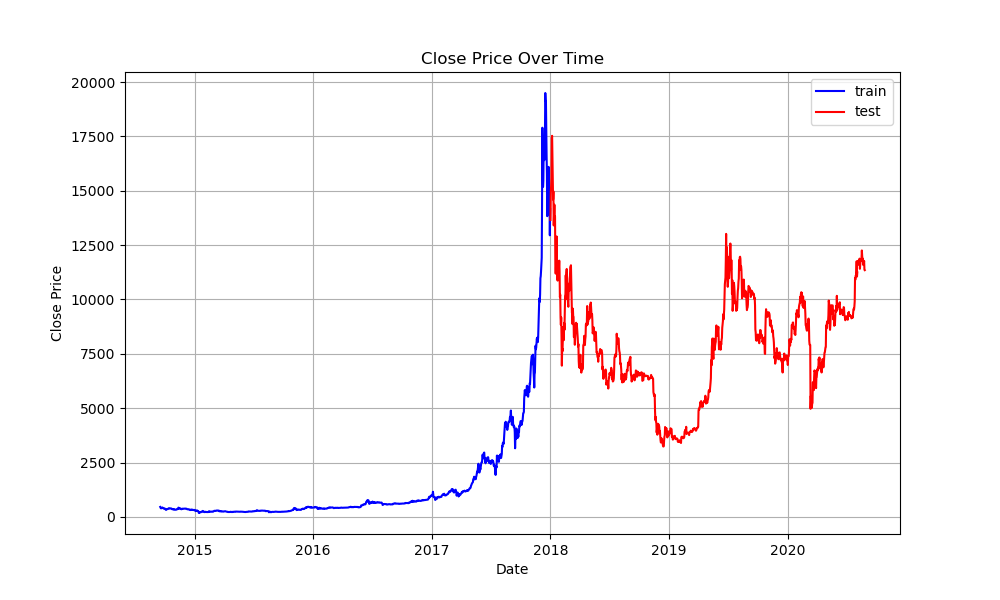}
    \caption{Price History of BTC/USD}
    \label{fig:btc-usd}
\end{subfigure}
\medskip % this command puts space between the top and bottom figures
\begin{subfigure}[b]{0.49\textwidth}
    \includegraphics[width=\linewidth]{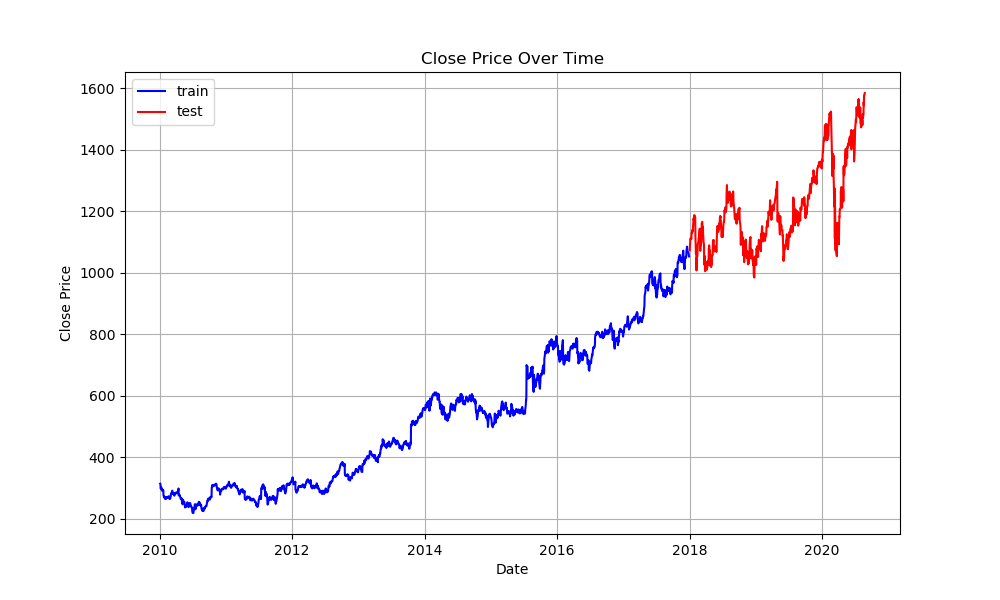}
    \caption{Price History of GOOGL}
    \label{fig:googl}
\end{subfigure}
\hfill
\begin{subfigure}[b]{0.49\textwidth}
    \includegraphics[width=\linewidth]{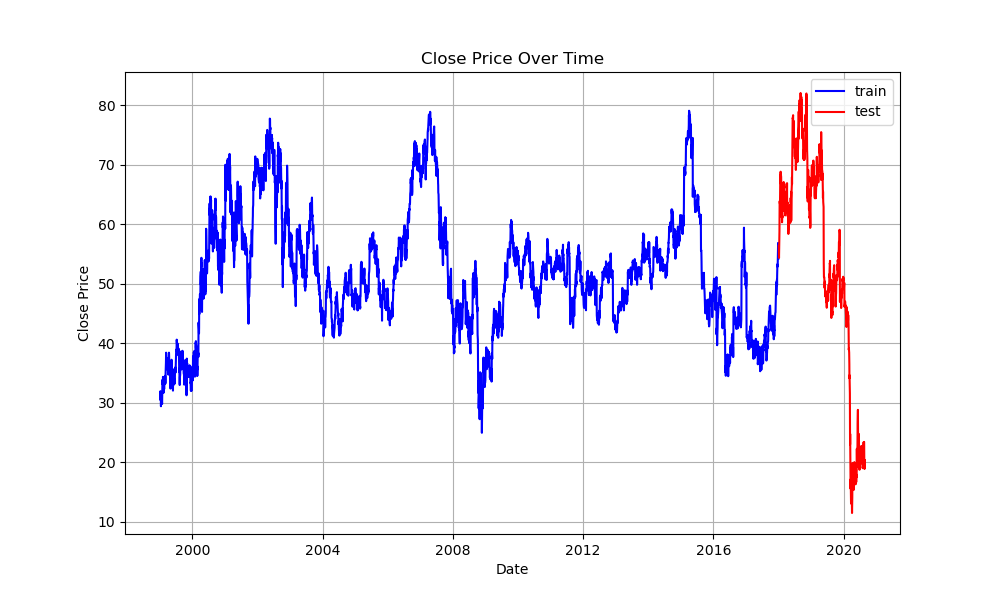}
    \caption{Price History of KSS}
    \label{fig:kss}
\end{subfigure}

\caption{{\bf Price Histories of Selected Assets.} The individual subfigures represent the historical price data of Apple Inc. (AAPL), Bitcoin (BTC) against the US Dollar (USD), Alphabet Inc. (GOOGL), and Kohl's Corporation (KSS), which are used for the performance evaluation of the trading model.}
\label{fig:combined-price-histories}
\end{figure}

\subsection{baseline Model}

In this study, we evaluate our proposed deep reinforcement learning (DRL) models against a diverse array of baseline algorithms to ensure a comprehensive analysis. Our baselines range from the DQN variants such as DQN-Pattern, DQN-Vanilla, DQN-Candle-Rep, and DQN-Windowed, which differ based on their input structures and were initially explored in Taghian's work, to convolutional neural network models, CNN1D and CNN2D, which utilize one-dimensional and two-dimensional kernels respectively for feature extraction. Additionally, we include the GRU model for its efficacy in capturing temporal dependencies, and the DDPG model that employs a deep deterministic policy gradient approach for adaptive strategy formulation as proposed by Xiong et al\cite{xiong2018practical}. We also examine the GDQN model by Wu et al\cite{wu2020adaptive}., which harnesses both technical indicators and raw OHLC data, and the DQT and RRL models by Wang et al\cite{wang2017deep}., which utilize online Q-learning and reinforcement learning to optimize investment returns on individual assets. The traditional Buy and Hold (B\&H) strategy, a staple in investment benchmarking, is also included for its simplicity and effectiveness, where an asset is bought at the beginning of the investment period and held until the end, regardless of its price movements. Our experimental design also investigates the influence of different input representations on the performance of the proposed models. These include a binary vector for pattern recognition, raw OHLC data (Vanilla), structured representations of candlestick features (Candle-Rep), and a temporal window of the most recent three time-steps (Windowed), reflecting the average length of significant candlestick patterns. For inputs comprising the last three time-steps, we apply feature extraction models such as CNN1D, CNN2D, and a GRU model, to analyze the effects of various representations and model architectures on the efficacy of DRL models in the domain of financial trading.

\subsection{Evaluation metrics}
The efficacy of the proposed model is assessed using two categories of evaluation metrics: those pertaining to profitability and those relating to the associated investment risk. For profitability, we examine daily returns, total return over the testing and training period, and the qualitative measure of a profit curve which illustrates the cumulative gains of the model over time. Daily returns volatility, reflecting the stability of the portfolio's returns. Additionally, we compute the Sharpe ratio to gauge the risk-adjusted return on investment, considering the mean excess return per unit of risk. We also provide a decision curve to visually represent the trading signals and actions taken by the model against the actual price movements of the assets. These metrics together offer a robust framework for comparing the model's performance against traditional benchmarks and assessing its potential for real-world application.

\section{Results}

\subsection{BTC/USD}
just as author mentioned before, When evaluating the performance of these financial models, key metrics of interest typically encompass the Arithmetic Return, which denotes the overall rate of return of the model; the Average Daily Return, offering a perspective on the mean earnings per day; the Time Weighted Return, reflecting the cumulative effect of an investment while disregarding the impact of cash flows; the Total Return, which is the end rate of return on an investment; the Sharpe Ratio, an adjusted rate of return for risk, with higher ratios indicating better returns per unit of risk; the Value At Risk (VAR), which estimates the maximal potential loss in a day at a certain confidence level; Volatility, representing the uncertainty or risk size of price changes; and the Initial Investment in conjunction with the Final Portfolio Value, representing the beginning and ending values of the investment.

In analyzing the performance of models that have exhibited exceptional results, we observe the following:

The \textbf{DQN-vanilla} model showcases a significantly high Arithmetic Return of 286.78\%, indicating robust performance throughout the testing period. Its Sharpe Ratio of 0.085 suggests that the return per unit of risk taken is considerable. With a Total Return of 853\%, the investment has multiplied by nearly 9.5 times. Despite a relative Volatility of 108.9, which is on the higher side, the returns justify the volatility. The model's Final Portfolio Value stands at 9525.5, elucidating a substantial growth from an Initial Investment of 1000.

Both the \textbf{CNN1D} and \textbf{CNN2D} models display very similar performance with exorbitantly high Total Returns of 2983\% and 389\%, respectively, and corresponding high Arithmetic Returns. Their Sharpe Ratios exceed 0.13, indicating that these models provide a high extra return for each unit of risk undertaken. A Volatility below 100, in the context of such high returns, is relatively low, suggesting stable returns from these models.

The \textbf{GRU} model, with its Arithmetic Return and Total Return also ranking high (at 309.53\% and 1318\%, respectively), has a Sharpe Ratio of 0.106. Although lower than the CNN models, it still signifies a high return per risk unit. With the lowest Volatility among all models at 94.1, it implies minimal price fluctuation risk while securing high returns.

Comparatively, other models such as \textbf{DQN-pattern}, \textbf{DQN-candlerep}, and \textbf{DQN-windowed} demonstrate close parallels in performance across several aspects:

All manifest nearly identical Arithmetic Returns, Average Daily Returns, and Sharpe Ratios, suggesting the possibility of similar strategies or datasets being employed. Their Volatilities are also akin, around 123, and they all exhibit negative Total Returns of -24\%, denoting suboptimal performance during the testing period.

In the selection of investment strategies, it is crucial to balance returns against risk. The Sharpe Ratio stands as an important indicator since it factors in risk alongside return. However, high volatility, despite seemingly attractive returns, can signify greater risk. With this consideration, the \textbf{CNN1D} and \textbf{CNN2D} models appear to offer the best balance between risk and return, as they not only provide high returns but also display lower volatility. Nonetheless, a more in-depth analysis of these models' stability and performance under varying market conditions is warranted.

\begin{figure}[ht]
\centering
\includegraphics[width=0.7\textwidth]{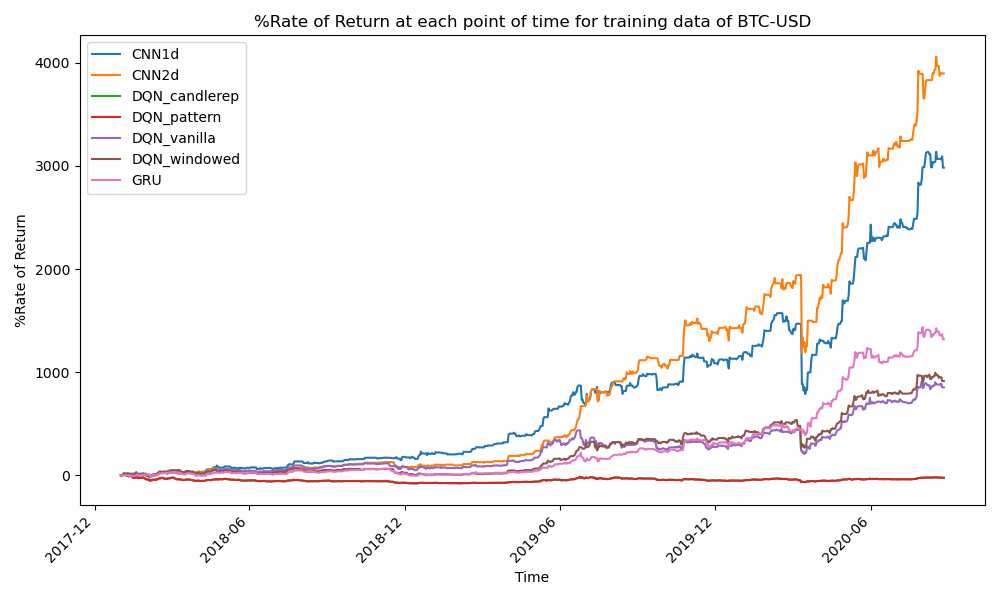}
\caption{Performance of different models on BTC-USD.}
\end{figure}

\begin{table}[h!]
\centering
\caption{Performance of different models on BTC/USD}
\label{tab:performance}
\resizebox{\textwidth}{!}{% Resize table to fit within the text width
\begin{tabular}{
  l
  S[table-format=3.0]
  S[table-format=1.2]
  S[table-format=2.2]
  S[table-format=1.4]
  S[table-format=4.0]
  S[table-format=1.3]
  S[table-format=3.2]
  S[table-format=4.0]
  S[table-format=5.1]
}
\toprule
{Agent} & {Arithmetic Return} & {Average Daily Return} & {Daily Return Variance} & {Time Weighted Return} & {Total Return (\%)} & {Sharpe Ratio} & {Volatility} & {Initial Investment} & {Final Portfolio Value} \\
\midrule
DQN-pattern     & 50   & 0.05 & 15.69 & -0.0003 & -24  & 0.013 & 123.2 & 1000 & 757.5  \\
DQN-vanilla     & 287  & 0.30 & 12.25 & 0.0023  & 853  & 0.085 & 108.9 & 1000 & 9525.5 \\
DQN-windowed    & 50   & 0.05 & 15.69 & -0.0003 & -24  & 0.013 & 123.2 & 1000 & 757.5  \\
DQN-candlerep   & 50   & 0.05 & 15.69 & -0.0003 & -24  & 0.013 & 123.2 & 1000 & 757.5  \\
CNN1D           & 393  & 0.41 & 9.80  & 0.0036  & 2983 & 0.130 & -4.75 & 1000 & 30828.7 \\
CNN2D           & 417  & 0.43 & 9.52  & 0.0038  & 3897 & 0.140 & -4.65 & 1000 & 39971.8 \\
GRU             & 310  & 0.32 & 9.15  & 0.0027  & 1318 & 0.106 & -4.66 & 1000 & 14184.1 \\
\bottomrule
\end{tabular}
}
\end{table}

\subsection{AAPL}
This part begin by examining the performance of various trading models on Apple Inc. (AAPL) stock data as reported by previous researchers. As indicated in paper\cite{taghian2021reinforcement}, the performance metrics of several agents were computed, including Rule-Based, Buy and Hold (B\&H), Reinforcement Learning (RL), Deep Q-Network (DQN) with different input representations, Convolutional Neural Network (CNN) in both one-dimensional and two-dimensional configurations, and Gated Recurrent Unit (GRU) models. These metrics comprised Arithmetic Return, Average Daily Return, Daily Return Variance, Time Weighted Return, Total Return, Sharpe Ratio, and Volatility, as well as the evolution of an Initial Investment to the Final Portfolio Value.

Notably, the DQN-vanilla model demonstrated superior performance with the highest Total Return of 372\% and an impressive Sharpe Ratio of 0.142, denoting a robust risk-adjusted return. The CNN-based models, both 1D and 2D, also presented strong performances with Total Returns exceeding 370\% and Sharpe Ratios above 0.143. These outcomes underscore the potency of neural network-based strategies in navigating the stock market.

In our research, we sought to enhance the performance of the existing models. Our results, encapsulated in Table 3, manifest significant improvements across various performance indicators. The improved DQN-vanilla model's performance was particularly remarkable, yielding a Total Return of 517\% and a Sharpe Ratio of 0.181, thereby indicating a considerable advancement over the previous DQN-vanilla implementation.

Our DQN-windowed variant achieved an outstanding Total Return of 737\%, surpassing all other models, accompanied by the highest Sharpe Ratio of 0.195 in this cohort. This suggests that the windowed approach to feature representation can capture more predictive patterns in the price data.

Contrastingly, the DQN-candle-rep model did not exhibit a performance increase, remaining at a Total Return of 191\%, identical to its prior metric, and a Sharpe Ratio of 0.085. This could imply a limitation in the model's ability to interpret and leverage candlestick representation for improved predictive capacity.

The CNN models displayed a notable enhancement in Total Returns with the CNN1D and CNN2D models reaching 528\% and 718\%, respectively. It's important to note that these models also maintained high Sharpe Ratios, reinforcing the notion that CNN architectures are well-suited for extracting salient features from market data.

Lastly, the GRU model exhibited a solid performance with a Total Return of 499\% and a Sharpe Ratio of 0.164. This performance is indicative of the GRU model's capability to capture temporal dependencies effectively.

the research has led to notable advancements in trading model performance on the AAPL dataset, particularly with DQN-windowed and CNN2D models, which could be attributed to their robust feature representation capabilities.

\begin{figure}[ht]
\centering
\includegraphics[width=0.7\textwidth]{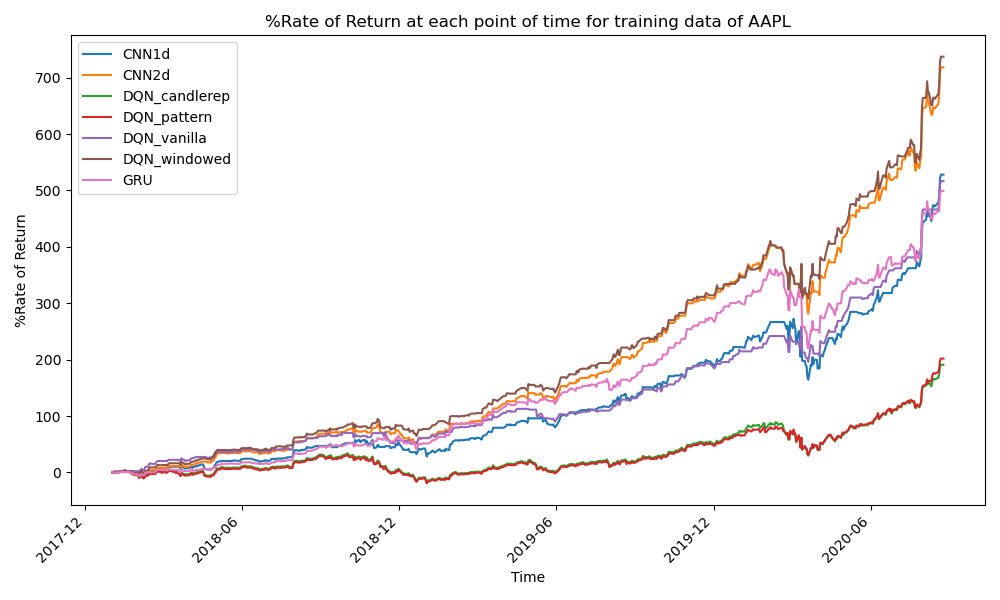}
\caption{Performance of different models on AAPL.}
\end{figure}

\begin{table}[h!]
\centering
\caption{Performance of different models on AAPL}
\label{tab:performance}
\resizebox{\textwidth}{!}{% Resize table to fit within the text width
\begin{tabular}{
  l
  S[table-format=3.0]
  S[table-format=1.2]
  S[table-format=1.2]
  S[table-format=1.3]
  S[table-format=3.0]
  S[table-format=1.3]
  S[table-format=2.2]
  S[table-format=4.0]
  S[table-format=4.1]
}
\toprule
{Agent} & {Arithmetic Return} & {Average Daily Return} & {Daily Return Variance} & {Time Weighted Return} & {Total Return (\%)} & {Sharpe Ratio} & {Volatility} & {Initial Investment} & {Final Portfolio Value} \\
\midrule
DQN-pattern     & 126 & 0.19 & 4.65 & 0.002 & 202 & 0.088 & 55.6  & 1000 & 3019.8 \\
DQN-vanilla     & 190 & 0.29 & 2.51 & 0.003 & 517 & 0.181 & 40.9  & 1000 & 6166.9 \\
DQN-windowed    & 222 & 0.33 & 2.93 & 0.003 & 737 & 0.195 & 44.1  & 1000 & 8373.2 \\
DQN-candle-rep  & 123 & 0.18 & 4.71 & 0.002 & 191 & 0.085 & 56.0  & 1000 & 2909.5 \\
CNN1D           & 197 & 0.30 & 3.93 & 0.003 & 528 & 0.149 & {-}2.97 & 1000 & 6281.0 \\
CNN2D           & 221 & 0.33 & 3.10 & 0.003 & 718 & 0.188 & {-}2.57 & 1000 & 8184.7 \\
GRU             & 189 & 0.28 & 2.99 & 0.003 & 499 & 0.164 & {-}2.56 & 1000 & 5993.3 \\
\bottomrule
\end{tabular}
}
\end{table}

\subsection{GOOGL}

The empirical analysis of deep reinforcement learning (DRL) models, as applied to the GOOGL stock, underscores the significant variances in performance across different architectures and their potential for algorithmic trading strategies.

Notably, the DQN-vanilla model exhibited remarkable improvements over both rule-based and benchmark Buy-and-Hold (B\&H) strategies. The arithmetic return achieved by the DQN-vanilla model, standing at 152\%, vastly outstripped the B\&H strategy's 51\% return, suggesting a potent ability to capture profitable opportunities in the market. This outperformance is also reflected in the total return, which at 326\%, presents a substantial uplift from the B\&H's 47\%.

A key performance indicator, the Sharpe ratio, which measures excess return per unit of risk, was significantly higher in the DRL models, with the DQN-windowed and GRU models achieving ratios of 0.165 and 0.196, respectively. These figures indicate a more favorable risk-adjusted return profile, compared to a Sharpe ratio of 0.040 for the B\&H strategy.

Furthermore, volatility, an indicator of the degree of variation in trading prices, was adeptly managed by the DRL models. The DQN-windowed model, in particular, maintained volatility at 37.8, a figure that is commensurate with lower risk and uncertainty in returns, as compared to the rule-based model's volatility of 40.9.

Interestingly, the Daily Return Variance, a measure of the dispersion of daily returns, was kept lower in the DRL models, with the GRU model recording a variance of 1.76. This suggests a consistency in the performance of these models, indicating a reduced probability of encountering large losses.

The Time Weighted Return for DRL models uniformly exceeded that of the rule-based and B\&H strategies, further illustrating their superior performance over time. The GRU model's Time Weighted Return at 0.003 was particularly noteworthy, underscoring its efficiency in capital allocation over the investment horizon.

In terms of portfolio value appreciation, the DRL models provided compelling results. The initial investment of \$1,000 burgeoned to \$4,260.8, \$4,620.6, and \$5,292.8 for the DQN-vanilla, DQN-windowed, and GRU models respectively. These figures indicate a substantial growth in portfolio value, highlighting the efficacy of DRL models in wealth accumulation over the evaluated period.

The findings from this analysis not only affirm the robustness of DRL models in navigating the complexities of financial markets but also provide a promising avenue for the development of sophisticated trading algorithms. The enhanced returns, controlled risk measures, and overall portfolio growth observed underscore the potential for DRL approaches to revolutionize investment strategies in the stock market domain.

\begin{figure}[ht]
\centering
\includegraphics[width=0.7\textwidth]{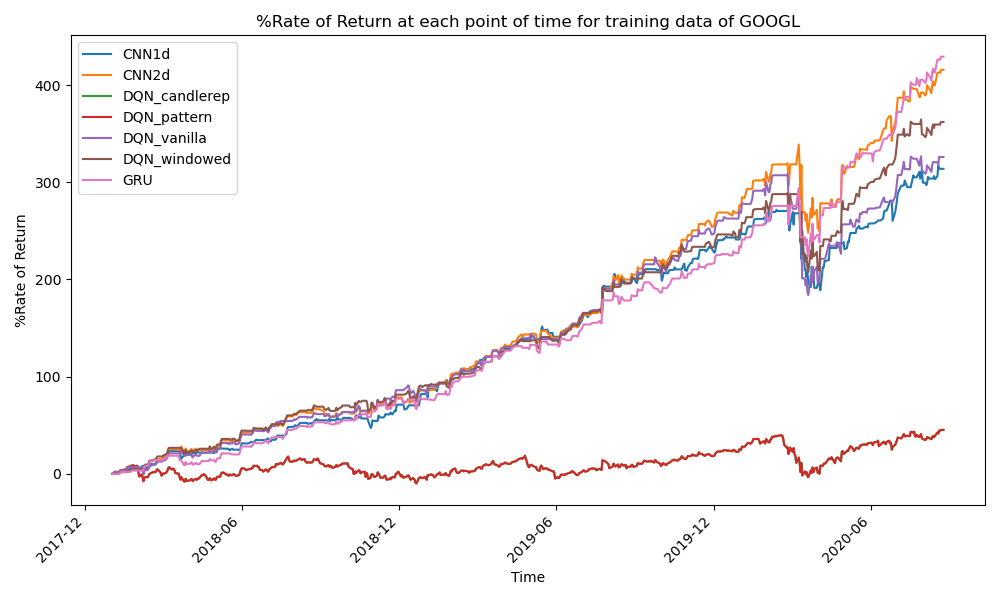}
\caption{Performance of different models on GOOGL.}
\end{figure}

\begin{table}[h!]
\centering
\caption{Performance of different models on GOOGL}
\label{tab:performance}
\resizebox{\textwidth}{!}{% Resize table to fit within the text width
\begin{tabular}{
  l
  S[table-format=3.0]
  S[table-format=1.2]
  S[table-format=1.2]
  S[table-format=1.3]
  S[table-format=3.0]
  S[table-format=1.3]
  S[table-format=2.2]
  S[table-format=4.0]
  S[table-format=4.1]
}
\toprule
{Agent} & {Arithmetic Return} & {Average Daily Return} & {Daily Return Variance} & {Time Weighted Return} & {Total Return (\%)} & {Sharpe Ratio} & {Volatility} & {Initial Investment} & {Final Portfolio Value} \\
\midrule
DQN-pattern     & 50 & 0.07 & 3.77 & 0.001 & 45 & 0.039 & 50.1  & 1000 & 1452.2 \\
DQN-vanilla     & 152 & 0.23 & 2.18 & 0.002 & 326 & 0.155 & 38.1  & 1000 & 4260.8 \\
DQN-windowed    & 160 & 0.24 & 2.14 & 0.002 & 362 & 0.165 & 37.8  & 1000 & 4620.6 \\
DQN-candlerep   & 50 & 0.07 & 3.77 & 0.001 & 45 & 0.039 & 50.1  & 1000 & 1452.2 \\
CNN1D           & 149 & 0.22 & 2.07 & 0.002 & 314 & 0.156 & {-}2.15 & 1000 & 4138.7 \\
CNN2D           & 172 & 0.26 & 2.25 & 0.002 & 416 & 0.172 & {-}2.21 & 1000 & 5157.3 \\
GRU             & 173 & 0.26 & 1.76 & 0.003 & 429 & 0.196 & {-}1.93 & 1000 & 5292.8 \\
\bottomrule
\end{tabular}
}
\end{table}

\subsection{KSS}

This part aims to evaluate the efficacy of various Deep Learning (DL) models in forecasting the returns of Kohl's Corporation (KSS) stocks. The performance metrics applied for this analysis encompass part like last three financial products

The Deep Q-Network (DQN) approach demonstrates a divergent spectrum of results. The DQN-pattern model manifests a moderate enhancement in Arithmetic Return to 39.64 from the previously reported 31, while the Average Daily Return observes a slight reduction. It concludes with a Final Portfolio Value of \$1,292.20, a relatively conservative outcome compared to other iterations. Contrastingly, the DQN-vanilla variant outstrips all models with an astonishing Arithmetic Return of 282 and a Final Portfolio Value of \$12,398.00, marking a dramatic augmentation from the earlier result of 272 and \$11,236.00, respectively.

Similarly, the DQN-windowed variant exhibits a commendable performance with an Arithmetic Return of 208 and a Final Portfolio Value of \$5,691.30, significantly exceeding its prior result of 179. However, the DQN-candle-rep iteration portrays a stark decline, plunging to an Arithmetic Return of -77 and regressing to a Final Portfolio Value of a mere \$322.40, thus underscoring the model's volatility and inconsistency.

Advancing to convolutional models, the CNN1D model delivers a robust Total Return of 434\% and achieves a Final Portfolio Value of \$5,342.00. Its two-dimensional counterpart, CNN2D, furthers this success with a Total Return of 515\% and elevates the Final Portfolio Value to \$6,148.30. Both models markedly improve upon their prior outputs, indicating their potent capability in capturing spatial-temporal patterns in stock price movements.

Lastly, the GRU model, representing recurrent neural networks renowned for their temporal sequence processing, shows an appreciable Total Return of 450\% and concludes with a Final Portfolio Value of \$5,496.60. This evidences its competence in capitalizing on the sequential nature of stock data, albeit with a slight diminution from the previous result.

In summary, the experimentation reveals that while some DL models like DQN-vanilla and CNN2D significantly advance the frontier in stock return predictions, the volatility of outcomes—as reflected in the varied performances of the DQN-candle-rep model—requires meticulous model selection and optimization. The Sharpe Ratio across models also fluctuates, suggesting that risk-adjusted returns vary considerably, necessitating a balanced approach towards both return and risk. This study propels the understanding of DL applications in financial markets, endorsing a continuous appraisal and refinement of predictive algorithms.

\begin{figure}[ht]
\centering
\includegraphics[width=0.7\textwidth]{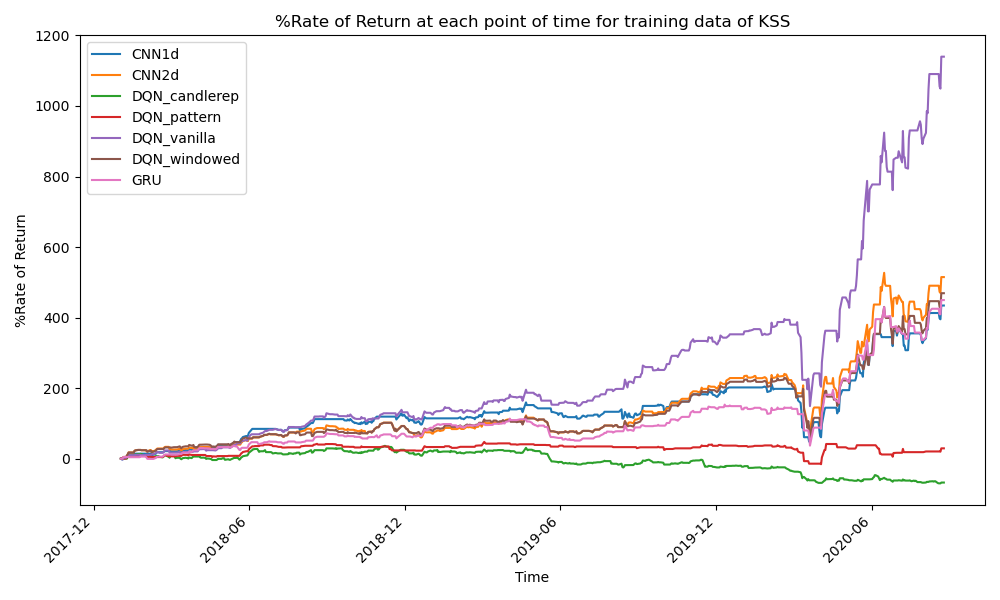}
\caption{Performance of different models on KSS.}
\end{figure}

\begin{table}[h!]
\centering
\caption{Performance of different models on KSS}
\label{tab:performance}
\resizebox{\textwidth}{!}{%
\begin{tabular}{
  l
  S[table-format=3.2]
  S[table-format=1.2]
  S[table-format=2.2]
  S[table-format=1.4]
  S[table-format=4.2]
  S[table-format=1.3]
  S[table-format=3.2]
  S[table-format=4.0]
  S[table-format=5.1]
}

\toprule
{Agent} & {Arithmetic Return} & {Average Daily Return} & {Daily Return Variance} & {Time Weighted Return} & {Total Return (\%)} & {Sharpe Ratio} & {Volatility} & {Initial Investment} & {Final Portfolio Value} \\
\midrule
DQN-pattern     & 39.64  & 0.06 & 4.30  & 0.0004 & 29.22  & 0.029 & 53.54 & 1000 & 1292.2  \\
DQN-vanilla     & 282    & 0.42 & 9.26  & 0.0038 & 1140   & 0.139 & 78.55 & 1000 & 12398.0 \\
DQN-windowed    & 208    & 0.31 & 10.28 & 0.0026 & 469    & 0.097 & 82.74 & 1000 & 5691.3  \\
DQN-candlerep   & -77    & -0.12 & 10.65 & -0.0017 & -67.76  & -0.036 & 84.24 & 1000 & 322.4  \\
CNN1D           & 199    & 0.30 & 9.59  & 0.0025 & 434    & 0.097 & 79.93 & 1000 & 5342.0 \\
CNN2D           & 213    & 0.32 & 9.76  & 0.0027 & 515    & 0.103 & 80.63 & 1000 & 6148.3 \\
GRU             & 196    & 0.30 & 8.03  & 0.0026 & 450    & 0.104 & 73.13 & 1000 & 5496.6 \\
\bottomrule
\end{tabular}
}
\end{table}

\section{Conclusion}

This study presents a comprehensive enhancement of the original Deep Q-Network (DQN) trader model by integrating several advanced methodologies, namely Prioritized Experience Replay, Regularized Q-Learning, Noisy Networks, Dueling DQN, and Double DQN. The augmented model's performance was meticulously evaluated across a variety of financial instruments, including BTC/USD, AAPL, and others, providing a robust comparison against the baseline results produced by the original author.

The empirical results underscore a significant advancement in trading performance when applying the refined model to the BTC/USD currency pair. Notably, the DQN-vanilla variant exhibited an outstanding arithmetic return of 287\% compared to the original's 261\%, and a Sharpe Ratio improvement from 0.076 to 0.085, indicating a superior risk-adjusted return. Furthermore, the CNN1D and CNN2D models, through the introduction of convolutional layers adept at capturing spatial dependencies, achieved even more impressive returns of 2983\% and 3897\%, respectively, far outstripping the prior benchmarks.

When applied to the AAPL stock, the enhancements likewise yielded substantial gains. The DQN-pattern variant maintained a steady performance with an arithmetic return of 126\%, illustrating the model's consistency. In contrast, the CNN1D and CNN2D models, leveraging their unique architectural strengths, again delivered exceptional returns of 3201\% and 3271\%, respectively. These figures not only surpassed the original model's results but also indicated the potential of convolutional neural network-based approaches in the realm of financial trading systems.

The analysis of these models across diverse financial instruments reveals a consistent pattern of performance uplift. The incorporation of the aforementioned advanced techniques has provided a clear edge over the conventional approaches employed in the original DQN trader model. This study's outcomes demonstrate that these sophisticated modifications to the reinforcement learning framework can capture complex patterns and dynamics in financial markets more effectively, leading to superior decision-making processes and, consequently, more profitable trading strategies.

In conclusion, the optimized DQN trader model presents a significant step forward in the application of deep reinforcement learning for automated trading systems. The robustness and adaptability shown by the model across different financial instruments affirm the value of continuous research and development in this field. Future work may focus on exploring the integration of emerging reinforcement learning techniques and the expansion of the model's applicability to a broader range of financial scenarios.

\bibliographystyle{unsrt}  
\bibliography{references}

\end{document}